\documentclass[journal]{IEEEtran}
\usepackage{amsmath}
\usepackage[]{siunitx}
\usepackage[]{units}
\usepackage{bm}
\usepackage{graphicx}

\ifCLASSINFOpdf
\else
\fi

\begin{document}
\title{Applying High-Resolution Visible Imagery to Satellite Melt Pond Fraction Retrieval: A Neural Network Approach}

\author{Qi Liu,
	Yawen Zhang,
	Qin Lv,
	Li Shang
	\thanks{Qi Liu and Yawen Zhang contributed equally to this work.}
	\thanks{Qi Liu and Li Shang are with the Department of Electrical, Computer and Energy Engineering,
		University of Colorado, Boulder, Colorado, 80309 USA}% <-this % stops a space
	\thanks{Yawen Zhang and Qin Lv are with the Department of Computer Science, University of Colorado, Boulder, Colorado, 80309 USA}% <-this % stops a space
}

\maketitle

\begin{abstract}
During summer, melt ponds have a significant influence on Arctic sea-ice albedo. The melt pond fraction (MPF) also has the ability to forecast the Arctic sea-ice in a certain period. It is important to retrieve accurate melt pond fraction (MPF) from satellite data for Arctic research. This paper proposes a satellite MPF retrieval model based on the multi-layer neural network, named MPF-NN. Our model uses multi-spectral satellite data as model input and MPF information from multi-site and multi-period visible imagery as prior knowledge for modeling. It can effectively model melt ponds evolution of different regions and periods over the Arctic. Evaluation results show that the MPF retrieved from MODIS data using the proposed model has an RMSE of 3.91\% and a correlation coefficient of 0.73. The seasonal distribution of MPF is also consistent with previous results. 
\end{abstract}


% \fi
\IEEEpeerreviewmaketitle

\section{Introduction}
\label{sec:intro}

\IEEEPARstart{M}{elt} ponds occur during the melting period of Arctic~\cite{fetterer1998observations, perovich2002seasonal}, influencing the sea ice albedo and solar energy partitioning on Arctic surface~\cite{perovich2012albedo, polashenski2012mechanisms}. The availability of an accurate Arctic melt pond fraction (MPF) dataset is crucial to explore the melt pond evolution~\cite{perovich2002aerial}. More importantly, it serves as an important parameter for Arctic climate study~\cite{flocco2010incorporation, flocco2012impact, schroder2014september}. Because of better coverage both in observation areas and period~\cite{tanaka2016estimation}, satellite remote sensing has become the main technique for Arctic MPF retrieval~\cite{tschudi2008, rosel2011comparison, kim2013melt, makynen2014estimation, han2016retrieval}. 

MPF retrieval algorithm plays an important role in producing a high-accuracy MPF dataset. Tschudi et al.~\cite{tschudi2008} proposed a spectral unmixing procedure to retrieve MPF in Beaufort/Chukchi Sea region. This algorithm is based on a set of linear equations within four classes (melt pond, open water, snow, and bare ice). The reflectivity of different classes was measured near Barrow, Alaska in June, 2004 and fixed a priori in the equations. The mean difference between the satellite retrieved MPF and aerial observed MPF is 1.5\%. Based on this algorithm, R{\"o}sel et al.~\cite{rosel2012melt} decreased it to three classes (melt pond, open water, and ice) while using similar equations to retrieve MPF. It should be noted that R{\"o}sel et al.~\cite{rosel2012melt} used the same fixed reflectivity as Tschudi et al.~\cite{tschudi2008} and expanded the MPF retrieval results to the whole Arctic. Their results were validated with National Snow and Ice Data Center (NSIDC) melt pond data~\cite{nsidcgt} at three sites, Beaufo for the Beaufort Sea, Canadian for the Canadian Arctic and Cafram for the Fram Strait. The root-mean-square error (RMSE) was 10.7\%  and R-squared was 0.28. Tanaka et al.~\cite{tanaka2016estimation} also used some prior information from ship-based MPF to build a linear model between AMSR-E brightness temperature (TB) and MPF. As a result, their model can only be applicable to areas with high sea ice concentrations and RMSE is 8.9\%. 

Melt ponds are highly variable both in time and space~\cite{polashenski2012mechanisms, webster2015seasonal}, which may result in substantial errors in satellite MPF retrieval~\cite{zege2015algorithm}. For MPF retrieval model, using fixed or inadequate prior knowledge for modeling could be problematic due to the lack of generality for temporal and spatial variations~\cite{tschudi2008}. A single site or period cannot represent the full range of melt pond variations, and using it as algorithm input could even impact the accuracy of retrieved MPF. 

To improve the overall accuracy of satellite MPF retrieval model, we concentrate on two aspects in this paper: 1) the development of the MPF retrieval algorithm, 2) prior knowledge used in the model. For the algorithm part, we propose a multi-layer neural network based MPF retrieval model, which investigates the potential of using neural networks for spectral information extraction and building the relationship between satellite spectral information and MPF. For the prior knowledge part, we gather the high-resolution visible images from different regions and periods, then extract MPF information as prior knowledge for the proposed model. Previously, these high-resolution visible images were only used as validation or comparison source for satellite retrieved MPF~\cite{rosel2012melt, zege2015algorithm, istomina2015meltpart1}. Our main contributions are summarized as follows.

\begin{itemize}
	\item The development of the MPF-NN model, a novel satellite MPF retrieval model based on multi-layer neural network, with prior knowledge extracted from multi-site and multi-period high-resolution visible images. 
	\item Demonstration of the model's effectiveness in automatic feature extraction from multi-spectral satellite data and retrieving high-accuracy MPF over the Arctic sea ice. 
\end{itemize}

Our paper is organized as follows. Section 2 presents the MPF-NN model and the data used in this work. Section 3 presents the results of model validation and retrieved MPF evolution over Arctic sea-ice. Section 4 discusses model observations and concludes this work.

\section{MPF-NN: A Neural Network based Satellite MPF Retrieval Model}
\label{sec:method}

\begin{figure}[h]
	\centering
	\includegraphics[width=0.48\textwidth]{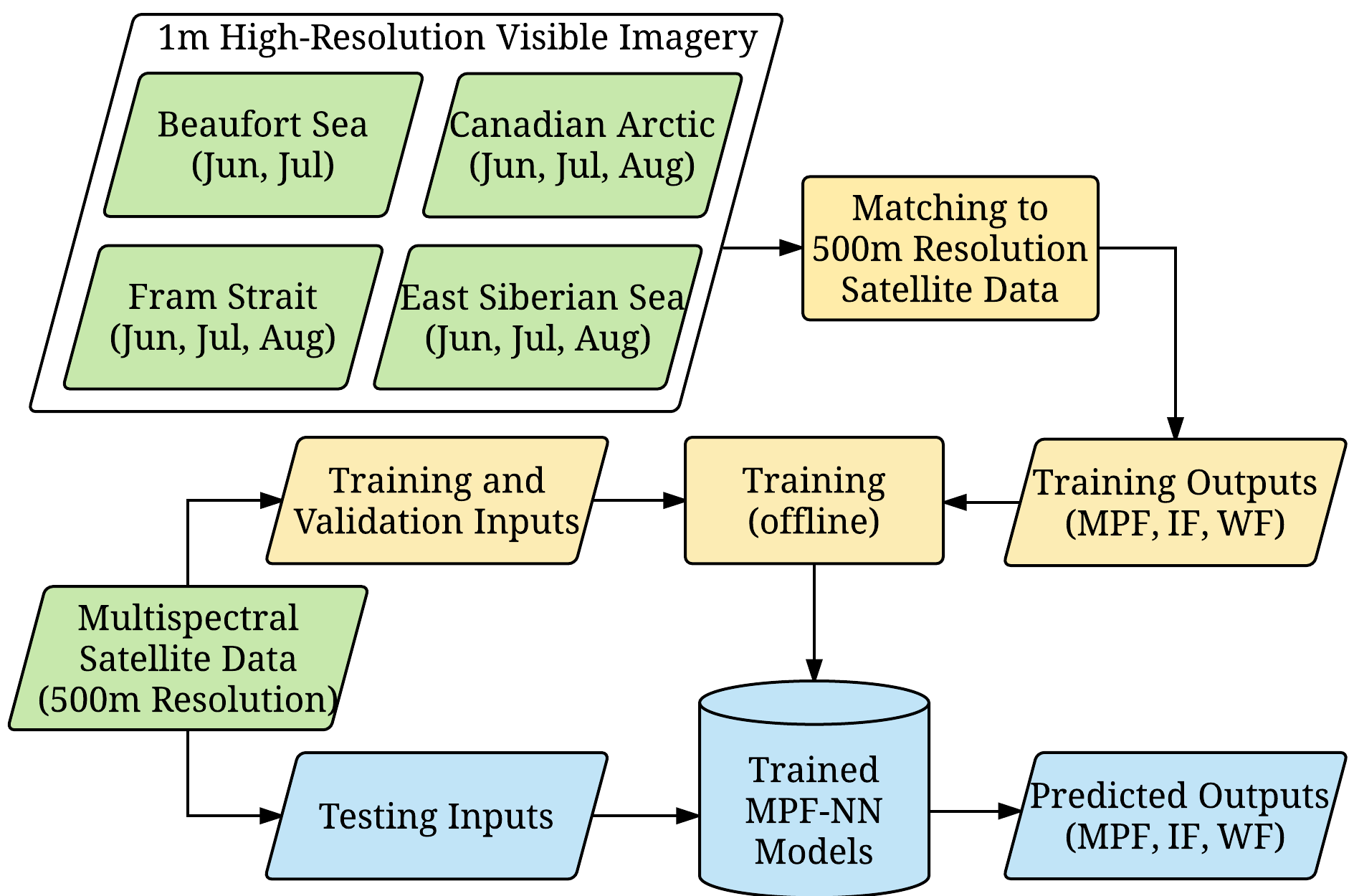}
	\caption{Design process of MPF-NN model.}
	\label{fig:methodchart}
	\vspace{-5mm}
\end{figure}

Our MPF-NN model is developed with the goal of 
accurately retrieving MPF at various spatial and temporal scales. The accuracy 
of our model relies on: (1) the completeness of spatial and temporal 
coverage provided by high-resolution visible images, which serve as prior knowledge 
and training data for our model; (2) the quality of input data that are used to 
derive MPF, e.g., the reflectance of multiple bands from satellite data. The flowchart 
in Fig.\ref{fig:methodchart} illustrates the development process of our MPF-NN model.

\begin{figure}[h!]
	\centering
	\includegraphics[width=0.25\textwidth]{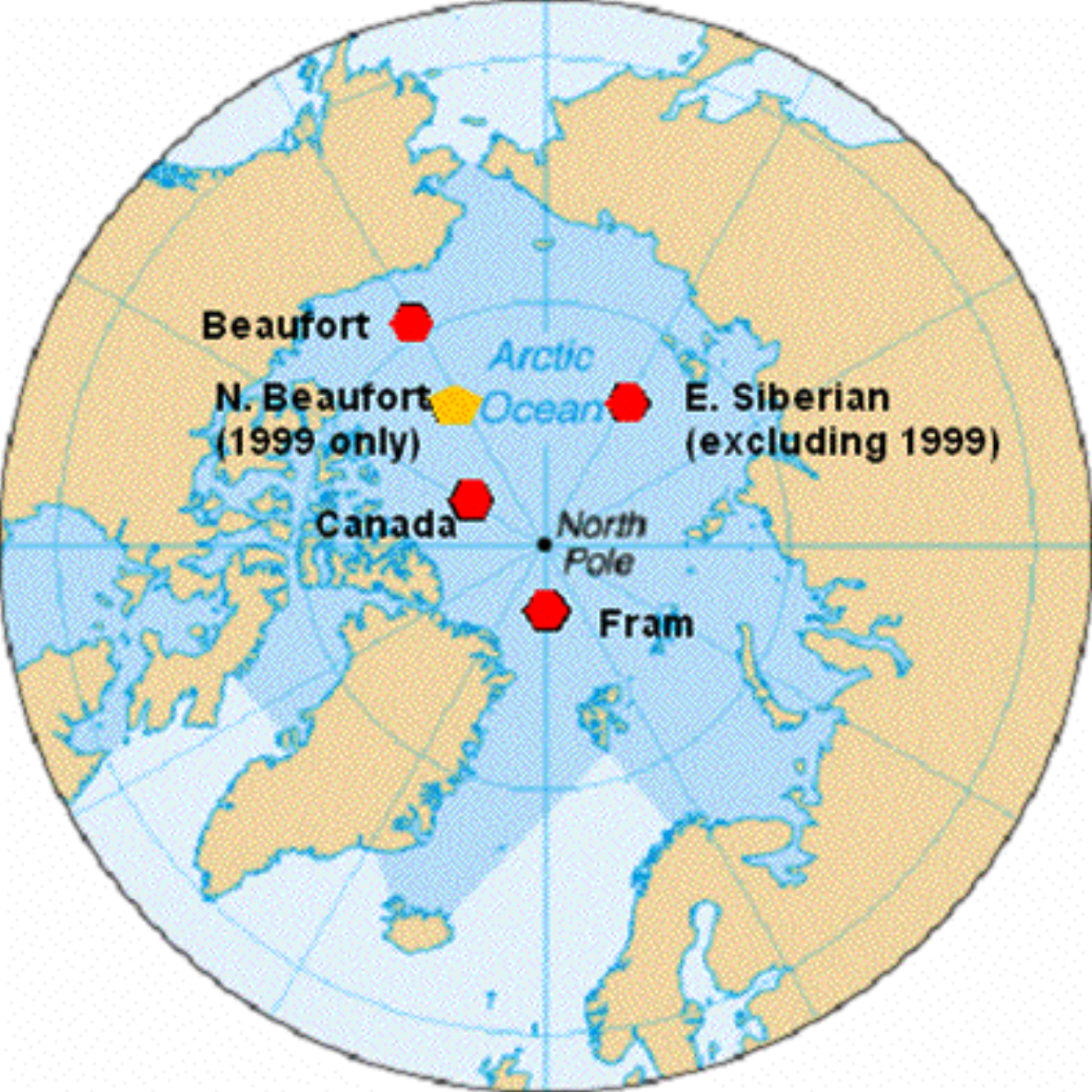}
	\caption{The Locations of four sites on Arctic sea (from NSIDC).}
	\label{fig:location}
	\vspace{-5mm}
\end{figure}

\begin{figure*}[h!]
	\centering
	\includegraphics[width=0.75\textwidth]{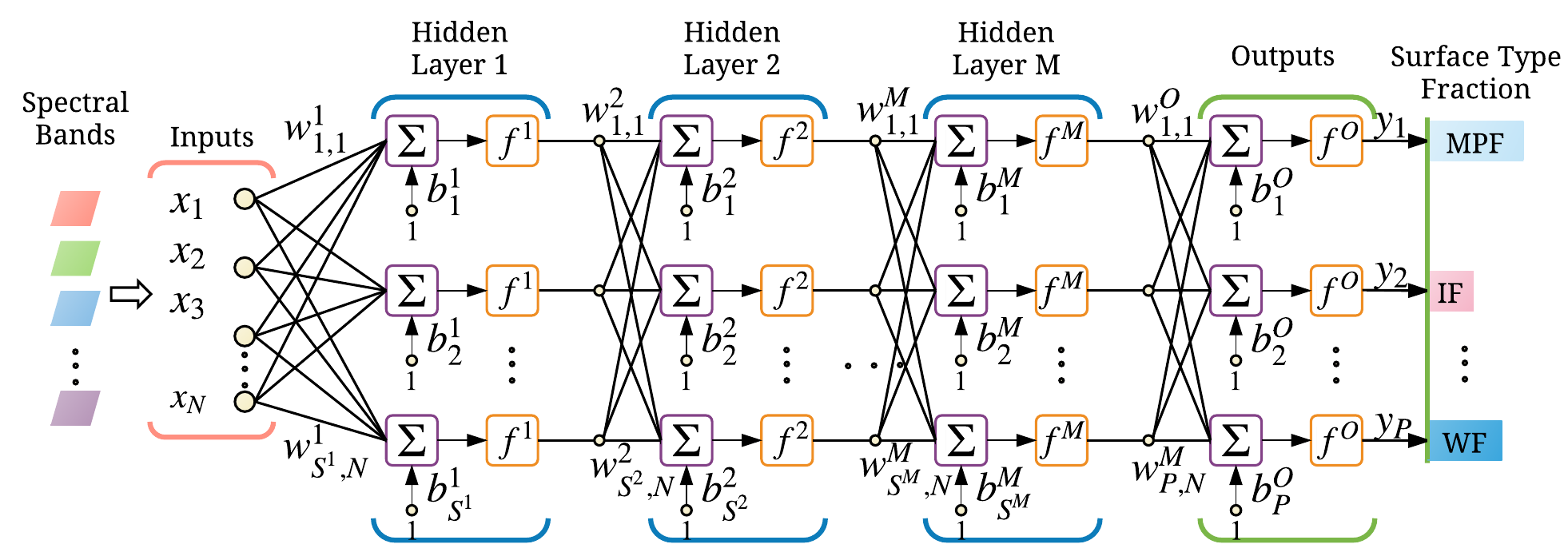}
	\caption{Illustration of a multi-layer neural network, which can be trained by a backward propagation learning algorithm through a series of layers. Inputs are reflectance from multi-spectral bands. Outputs are the fractions of multiple surface types, e.g., MPF, IF and WF.}
	\label{fig:nn}
	\vspace{-3mm}
\end{figure*}

\subsection{Data Source for MPF-NN}

\subsubsection{Input Data}

Our model utilizes multispectral satellite data as inputs. We select MODIS 500\si{m} daily surface reflectance products (MOD09GA) and use reflectance band 1 to 7 as model inputs. The wavelength of band 1-7 are 620-670\si{nm}, 841-876\si{nm}, 459-479\si{nm}, 545-565\si{nm}, 1230-1250\si{nm}, 1628-1652\si{nm}, and 2105-2155\si{nm}, respectively. All the cloud pixels in the daily product are removed. 
%MOD09GA embedded "state-1\si{km}" layer is extracted for cloud mask. 

\subsubsection{Output Data}

Our MPF-NN has three outputs: melt pond fraction (MPF), ice fraction (IF) and open water fraction (WF), which are derived from NSIDC melt pond imagery~\cite{nsidcgt}. The NSIDC dataset came from a collection of high-resolution visible imagery, which was acquired over different Arctic sites during the summers of 1999, 2000, and 2001.
%Cloud pixels were manually masked. 
% image resolution is 1\si{m} and each covers about a square area of 10\si{km^{2}}. 
%Each image has two (water and ice) or three (pond, open water, and ice) surface types.
%To reduce the uncertainty in this dataset, we also manually remove the images with obvious misclassification.

In this study, high-resolution images from four sites: Beaufort Sea (73$^{\circ}$N, 150$^{\circ}$W), Canadian Arctic (85$^{\circ}$N, 120$^{\circ}$W), Fram Strait (85$^{\circ}$N, 0$^{\circ}$E) and East Siberian Sea (82$^{\circ}$N, 150$^{\circ}$E) in the summers of 2000 and 2001 are gathered as shown in Fig.\ref{fig:location}. In order to pair the input and output datasets, we re-project NSIDC imagery data to the Sinusoidal Projection (same as MOD09GA), down-sample the 1\si{m} resolution grids to 500\si{m} grids by averaging pixels values within a $500 \times 500$\si{m} grid and calculate the MPF, IF and WF in each grid corresponding to the grid of MOD09GA. In total, we train the model with 3,925 pairs of input and output data. 

\subsection{MPF-NN Architecture}

Our MPF-NN utilizes a multi-layer neural network architecture, which contains multiple hidden layers of units between the input and output layers~\cite{lecun2015deep}, and is capable of extracting sophisticated features from input data and modeling non-linear input/output relationships~\cite{hornik1991approximation,nn1}. As illustrated in Fig.~\ref{fig:nn}, the network has one input layer, $M$ hidden layers,  
and one output layer. The input layer consists of $N$ units, each hidden layer consists 
of $S^{M}$ units and the output layer contains $P$ units. A hyperbolic tangent (tanh) 
activation function is used after each hidden layer to capture the nonlinear relationship 
between the multi-spectral bands and the fractions of different surface classes. 
During model training and validating, backward propagation learning algorithm and 
adaptive moment estimation (ADAM) method are applied for optimizing the weights 
$\bm{w}$ and bias $\bm{b}$ before and after each hidden layer~\cite{kingma2014adam}. The nonlinear relationship between input $\bm{x}$ and $\bm{y}$ is represented in Eq.\ref{eq:nn}, where $f$ is an activation function after each hidden layer and output layer.
\begin{equation}
\bm{y} = f^{O}(\bm{w}^{M}\cdot \cdot \cdot f^{2}(\bm{w}^{2}f^{1}(\bm{w}^{1}\bm{x}+\bm{b}^{1}))+\bm{b}^{2})+\cdot \cdot \cdot \bm{b}^{M}))))
\label{eq:nn}
\end{equation}

%Although multi-layer neural network is a powerful data modeling approach, there is no formally 
%established theory regarding the selection of the optimal number of hidden layers 
%and number of hidden units. 
Here, we briefly describe the design of our MPF-NN 
with an example of using MOD09GA data. We use a neural network with 
3 hidden layers, each layer consisting of 7, 10, 10 units, respectively. The rationale 
behind this is: (1) we choose the number of units for each layer that is comparable 
with the number of input bands. For instance, all 7 bands of MOD09GA data are 
used as inputs, so we choose 7 or more units for each layer. (2) for the number of 
layers, we start from the simplest structure, i.e., 2 layers, then keep adding layers 
until the performance of the model stops improving. However, this empirical approach 
can lead to over-fitting with the incrementation of model depth, especially, when 
there are limited training data. In order to prevent over-fitting, each hidden layer is 
followed by a dropout layer. Dropout is a regularization technique to prevent over-fitting, which consists of randomly setting a fraction $p$ of hidden layer units to 0 at each 
update during training~\cite{srivastava2014dropout}. The typical range of $p$ is 
from 0.2 to 0.5. Empirically, we use 0.2 for the second and third hidden layers.

\section{Results}
\label{sec:results}
We first validate the performance of our MPF-NN model on feature extraction and MPF retrieval accuracy. The results are compared with the MPF produced by the CICE hindcasts melt pond fraction model~\cite{schroder2014september}. Finally, we analyze the retrieved MPF over the entire Arctic sea ice. 

\subsection{Model Validation}

\subsubsection{Feature Extraction}

\begin{figure}[h]
	\centering
	\includegraphics[width=0.41\textwidth]{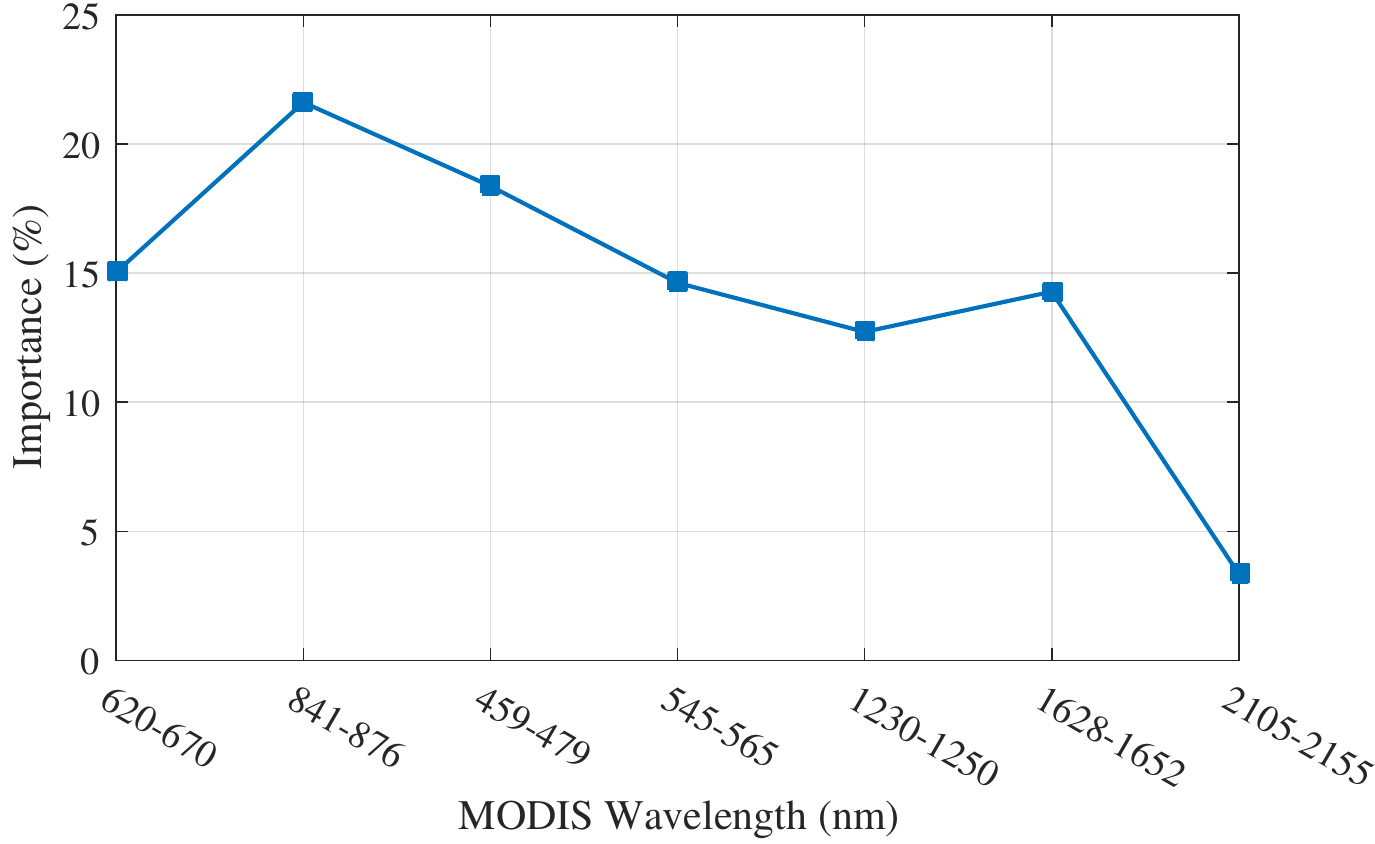}
	\caption{Importance (\%) score of each band in differentiating melt ponds from ice and open water.}
	\label{fig:featurerank}
	\vspace{-3mm}
\end{figure}

Typically, satellite imagery has multiple bands and hence it is essential to interpret how much each band contributes to the output, i.e., which bands are more important in differentiating melt ponds from ice and open water. Band 1, 2, 3 are utilized in the spectral unmixing algorithm~\cite{rosel2012melt,tschudi2008}. In our MPF-NN model, all 7 bands from MOD09GA product are used to further improve the modeling accuracy. Here, we use MPF-NN to extract the importance of different bands automatically. Fig.~\ref{fig:featurerank} shows the accumulated importance of each band to the output, which is computed by Eq.\ref{eq:fearank}. $\bm{r}$ is the importance vector and it has $k$ elements. Each element $r_{k}$ represents the importance of the $k$-th band, $\bm{w}$ represents the input unit weights extracted from each hidden layer.
\begin{equation}
\bm{r} = \frac{\sum |\bm{w}^{1}||\bm{w}^{2}|\cdot \cdot \cdot|\bm{w}^{M}|}{\sum\sum |\bm{w}^{1}||\bm{w}^{2}|\cdot \cdot \cdot|\bm{w}^{M}|}
\label{eq:fearank}
\end{equation}
The importance scores of band 1-7 are shown in Fig.\ref{fig:featurerank}. We can see 
that bands 1-6 contribute most to the final output with importance scores larger than 10\%. Band 2 contributes the most and band 7 contributes the least. Our feature interpretation result is consistent with band selection in the spectral unmixing algorithm~\cite{rosel2012melt} as bands 1-3 show the highest importance scores among the 7 bands. However, our result indicates that besides bands 1-3, bands 4-6 can also provide additional information for melt ponds, ice and open water separation, together covering around 40\% of the total contribution to the final output. 

\begin{figure}[h]
	\centering
	\includegraphics[width=0.43\textwidth]{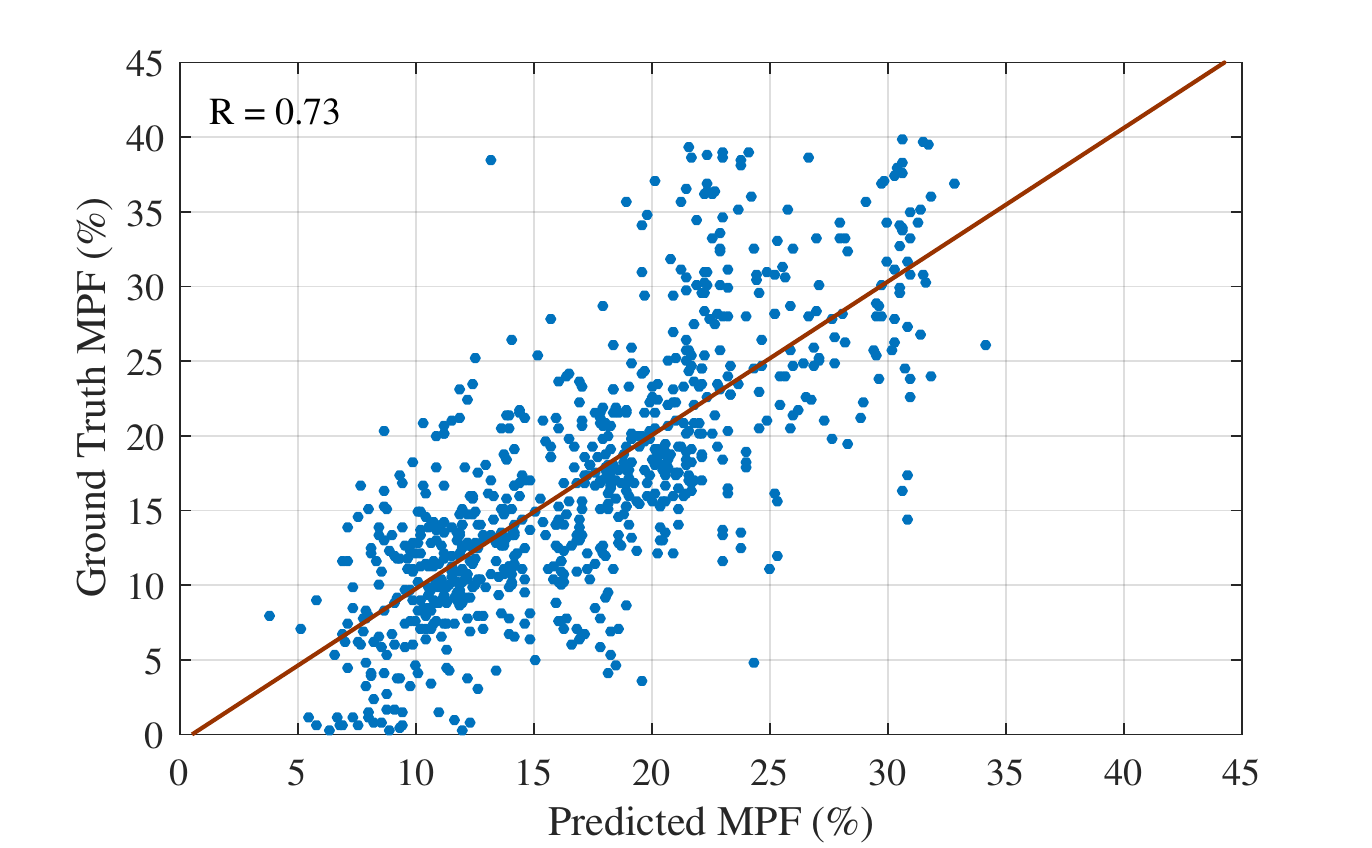}
	\caption{Scatter plot showing the MPF retrieved by our MPF-NN model and 
		the ground truth MPF derived from NSIDC dataset.}
	\label{fig:correlation_gt}
	\vspace{-3mm}
\end{figure}

%A linear function has been 
%fit to the data (red line). Pearson's correlation coefficient is $R = 0.73$

\subsubsection{Model Accuracy}

A total of 3,925 grids ($500 \times 500$\si{m}) are selected by matching the high-resolution visible images and MODIS data. 70\% of them are used for training the model, 10\% are used as validation and 20\% are used as the test dataset. A 5-fold 
cross-validation is performed by randomly swapping grids from the test dataset into the training and validation datasets. The model's training process converges in 10,000 epochs. Fig.~\ref{fig:correlation_gt} shows the correlation between MPF retrieved from the trained MPF-NN model and ground truth melt pond fraction derived from NSIDC high-resolution imagery. Overall, our MPF-NN model achieves a RMSE of 3.91\% and a correlation coefficient of 0.73. 

\begin{table*}[]
	\centering
	\caption{Sites MPF Statistics}
	\label{tbpeaks}
	\begin{tabular}{l|cccc}
		\hline
		\multicolumn{1}{c|}{\textbf{Model}}      & \multicolumn{4}{c}{\textbf{MPF-NN}}                                                                                                                                  \\ \hline
		\multicolumn{1}{c|}{\textbf{Statistics}} & \multicolumn{1}{l}{\textbf{Beaufort}} & \multicolumn{1}{l}{\textbf{Canadian}} & \multicolumn{1}{l}{\textbf{East Siberian}} & \multicolumn{1}{l}{\textbf{Fram Strait}} \\ \hline
		\textbf{Date (1st Peak)}                 & 06-27                                 & 07-03                                 & 07-06                                      & 07-12                                    \\
		\textbf{Data (2nd Peak)}                 & 07-23                                 & 07-23                                 & 07-24                                      & 07-23                                    \\
		\textbf{Mean MPF (\%)}                   & 18.22                                 & 11.24                                 & 13.04                                      & 11.56                                    \\ \hline
	\end{tabular}
\end{table*}

\begin{figure*}[h!]
	\centering
	\includegraphics[width=0.345\textwidth]{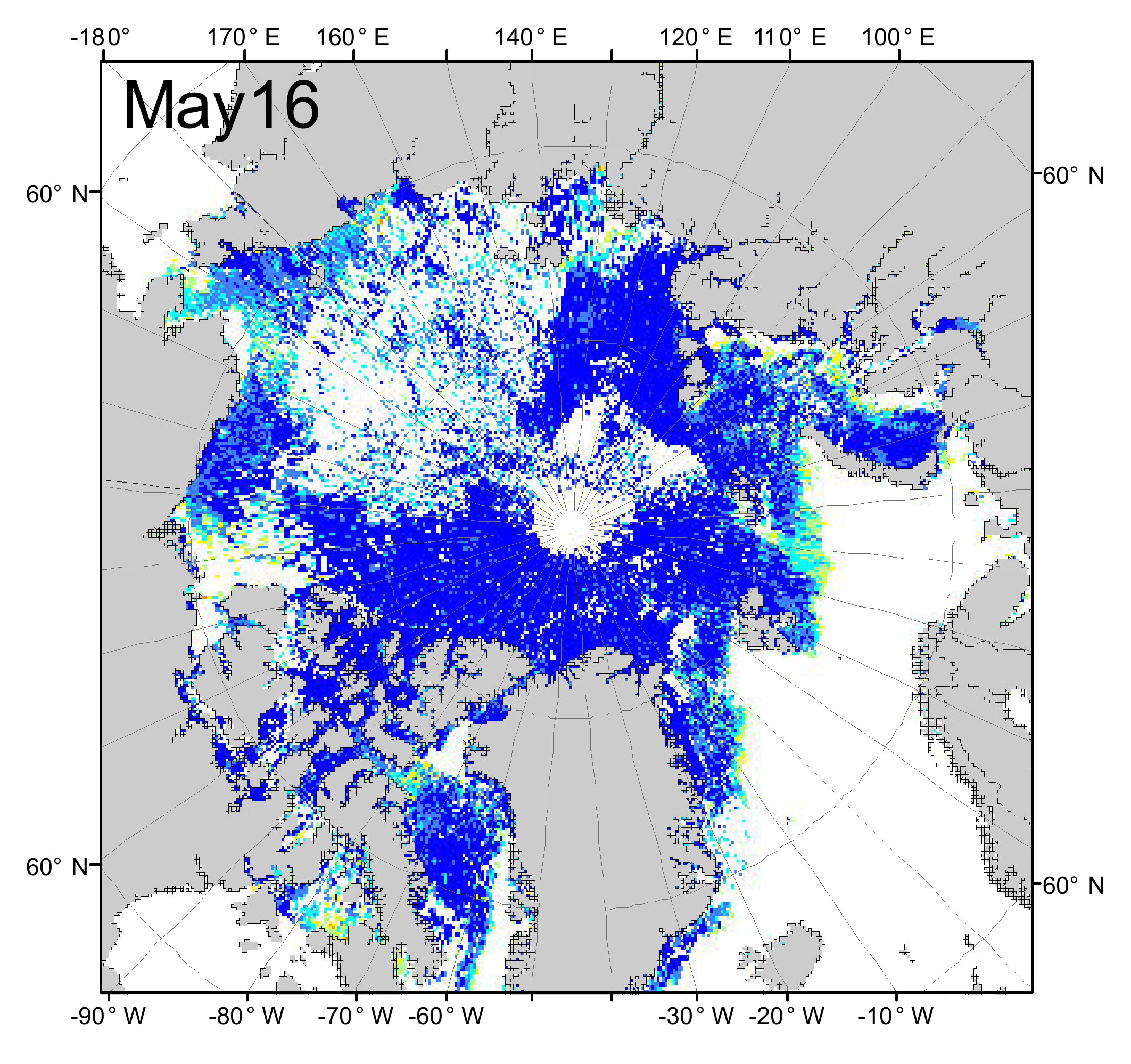}
	\includegraphics[width=0.345\textwidth]{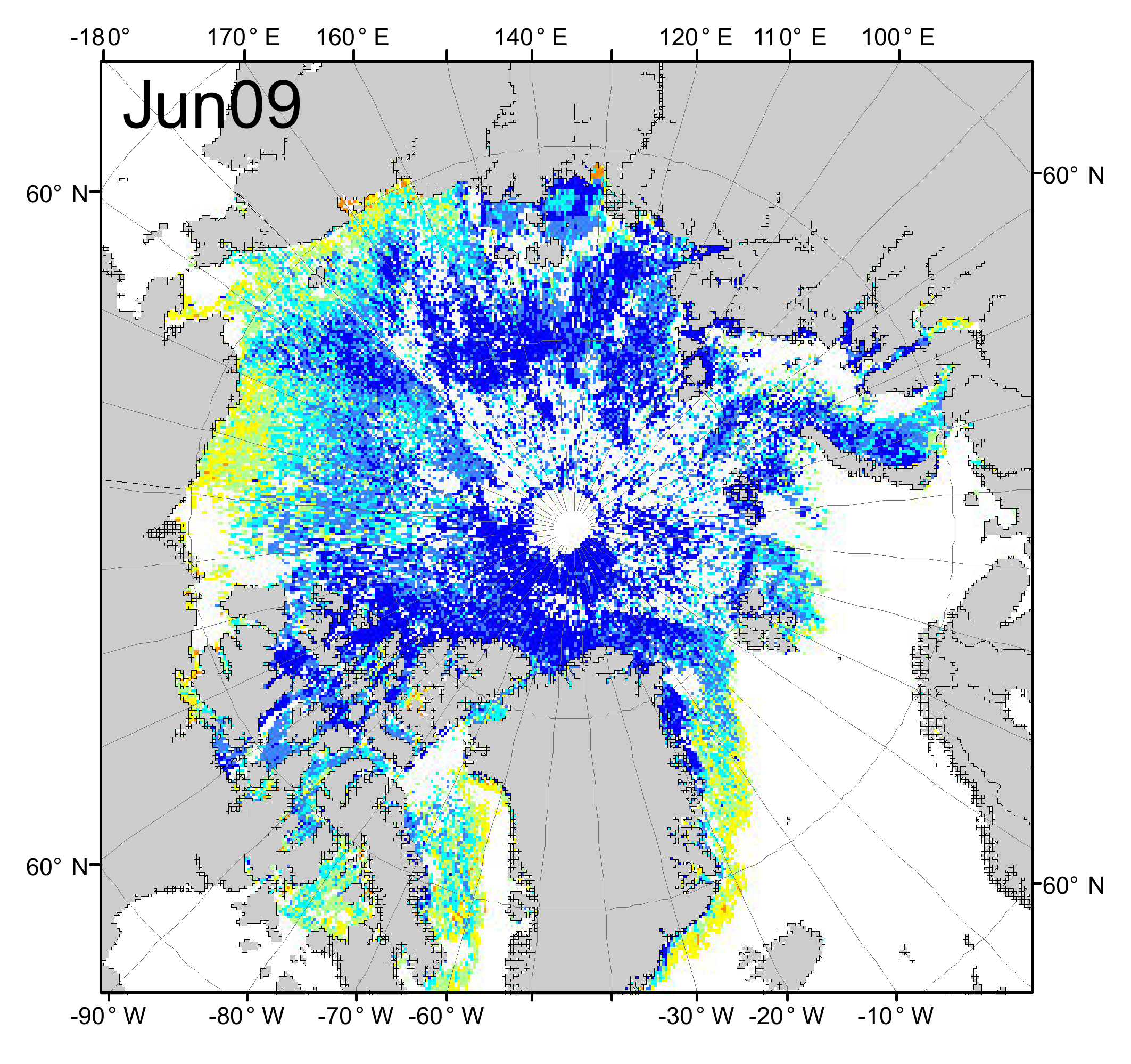}\\
	\includegraphics[width=0.345\textwidth]{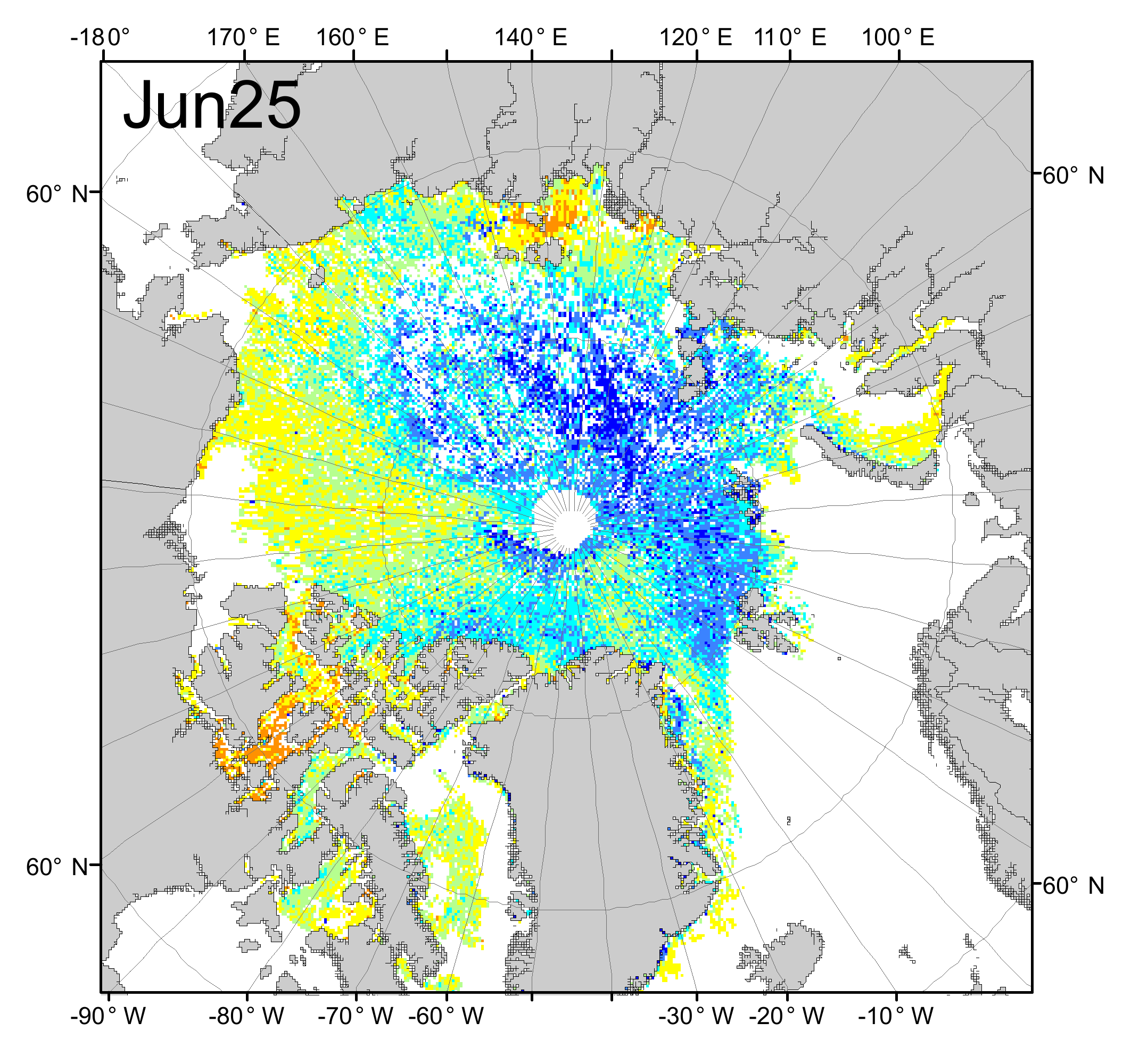}
	\includegraphics[width=0.345\textwidth]{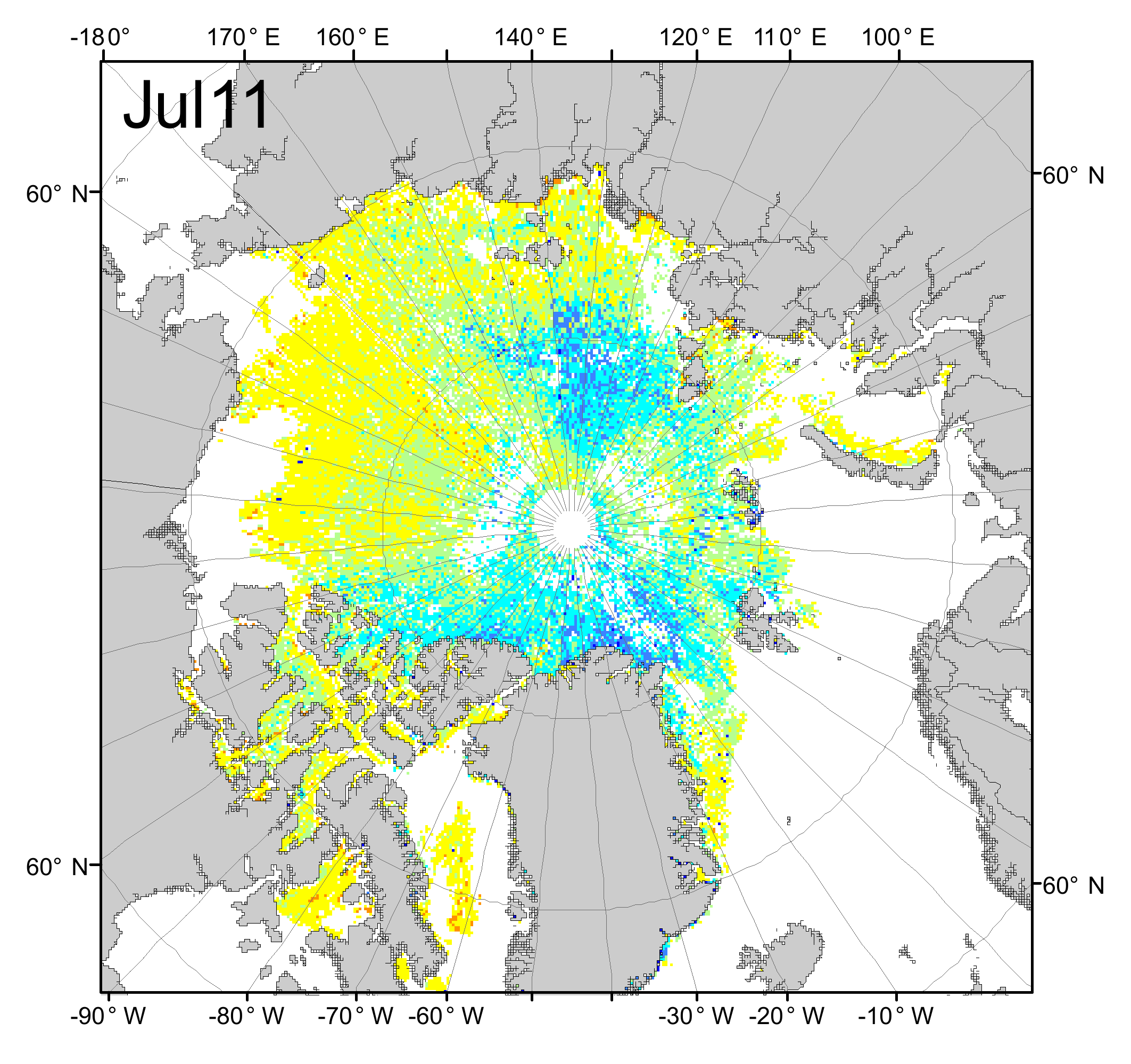}\\
	\includegraphics[width=0.345\textwidth]{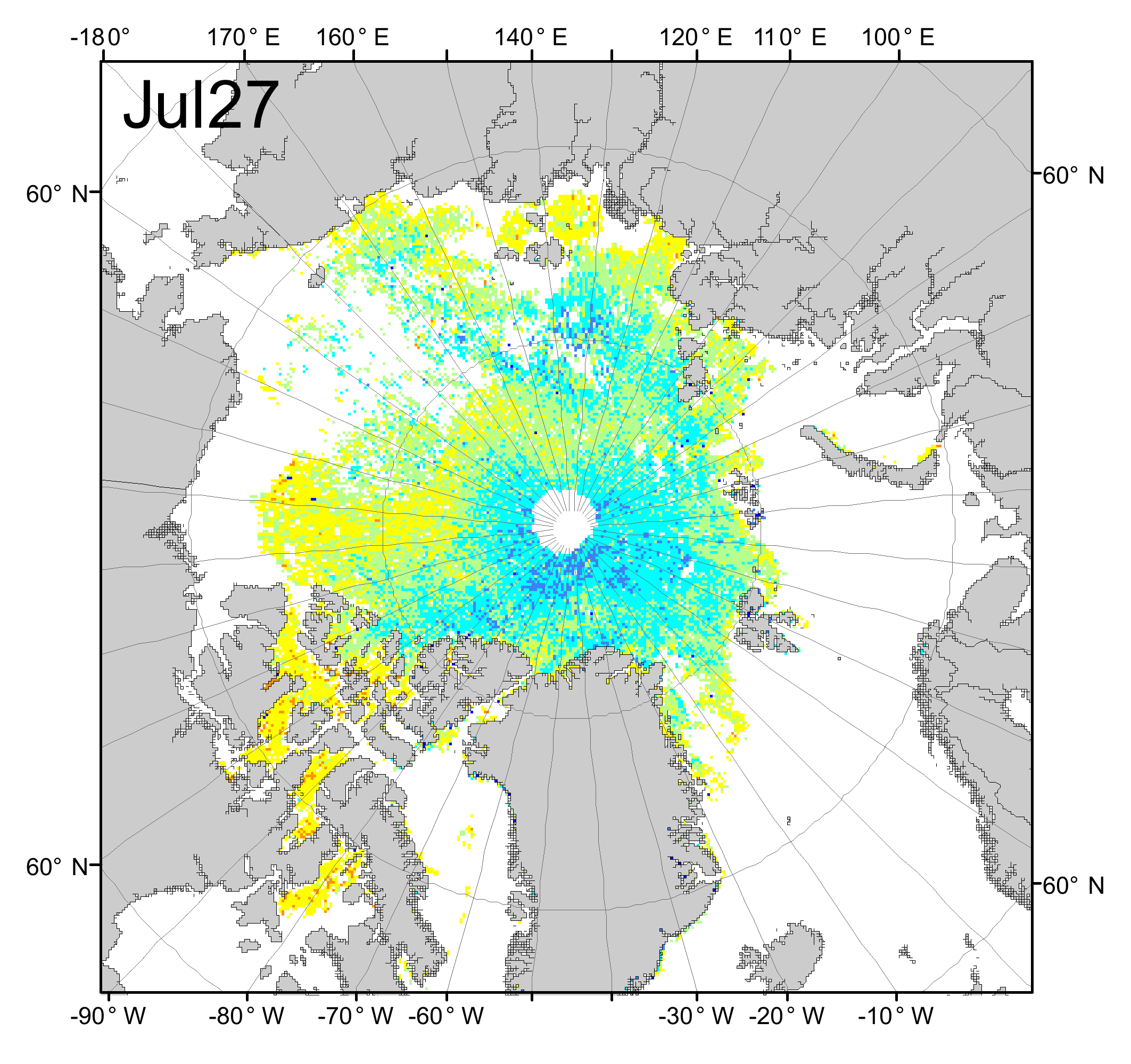}
	\includegraphics[width=0.345\textwidth]{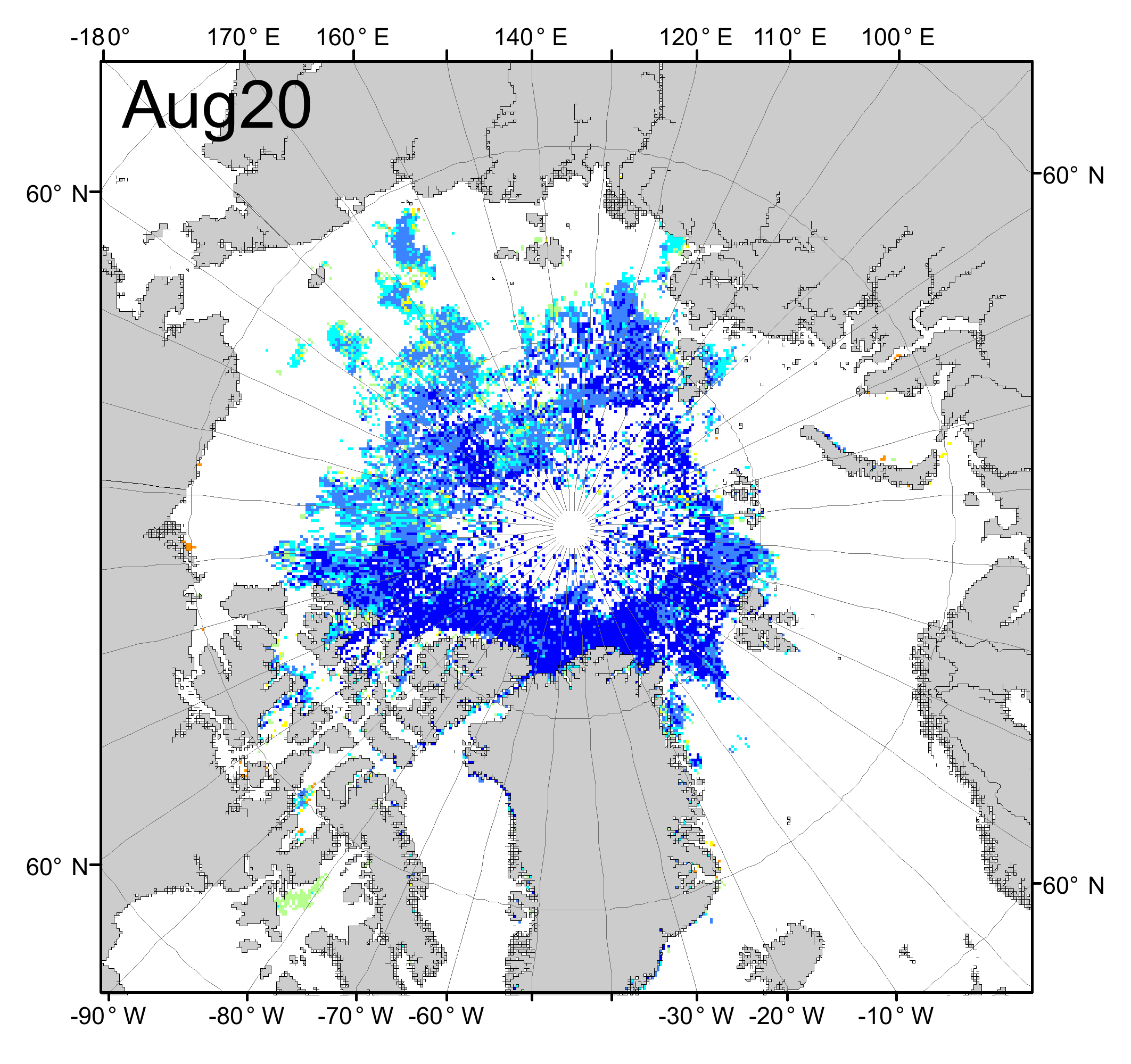}
	\includegraphics[width=0.6\textwidth]{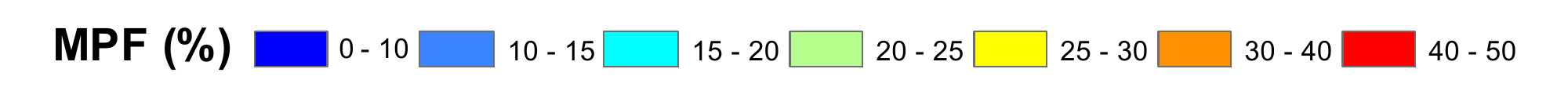}
	\caption{Arctic MPF map in the summer of 2008.}
	\label{fig:Arctic_MPF_Map}
	\vspace{-3mm}
\end{figure*}

\subsection{Melt Pond Fraction Evolution}
Our model is applied to retrieve the MPF at different sites (Beaufort Sea, Canadian Arctic, Fram Strait) and for the whole Arctic. The evolution analysis can illustrate the temporal and spatial variation of melt ponds. 

\subsubsection{Four Sites}

For each site, we select a 10\si{km^{2}} square area surrounding it to calculate its MPF. Table.~\ref{tbpeaks} shows several key statistics such as first peak, second peak and mean MPF. The mean MPF is calculated by averaging the MPF from May to the first week of September except for Beaufort Sea site, where the time period spans from May to the end of July. As it shows, Beaufort Sea has the highest mean MPF (18.22\%), while East Siberian has the second highest mean MPF (13.04\%), which is slightly higher than the mean MPF in Canadian Arctic and Fram Strait. This result is highly related to the geographical latitudes where those sites are located, as Beaufort Sea has lower latitude than the other three sites. 

For occurrence of peaks, we find that the two peaks for Canadian Arctic, East Siberian, and Fram Strait derived from MPF-NN model share similar characteristics with each other. For instance, as shown in Table.~\ref{tbpeaks}, these three sites have the first peak in early July and the second peak in late July. For Beaufort sea, it reaches the first peak earlier than the other three sites. We also compare our peak result with the Arctic MPF peaks produced by the CICE hindcasts model~\cite{schroder2014september}, which has the first peak at July 10th and the second peak at July 16th. Our peak results agree with Arctic overall peaks. 

Our MPF-NN model utilizes melt ponds information from multiple sites and multiple time periods. Therefore, the MPF-NN model has the capability to retrieve MPF at each individual site and effectively shows the spatial-temporal variability of MPF trend. 
%
%Additionally, although the CICE hindcasts model and the spectral unmixing model both produce MPF trend over the entire Arctic sea ice, the predictions are very different in terms of the peak formation periods.

\subsubsection{Arctic}
We further evaluate the robustness of the MPF-NN model by retreving MPF in a different year when the ground truth data for model training is not available. Because of cloud coverage, we select MODIS 500\si{m} 8-day surface reflectance products (MOD09A1) instead of daily products to produce the Arctic MPF map during the summer of 2008. The date showing on the figure is the start date of this 8-day period. All cloud pixels are removed. As shown in Fig.~\ref{fig:Arctic_MPF_Map}, during summer, the general MPF evolution characteristics retrieved by our model is consistent with other models~\cite{rosel2012melt, zege2015algorithm}, which strongly increases in June and July, reaching a value of 20-30\%. The minimum MPF appears in May and the end of August which are less than 10\%. Also, we notice that melt ponds start later at higher latitude, which agrees with the finding of Zege, E et al.~\cite{zege2015algorithm}. 

\section{Conclusions}
\label{sec:disc}
In this work, we developed MPF-NN, a novel multi-layer neural network based approach to retrieve MPF from satellite data. MPF-NN utilizes multi-site and multi-period high-resolution visible imagery as prior knowledge and hence has a better capability to capture the spatial-temporal variability of melt ponds than single field observation. An automated method was designed to interpret the contribution of different spectral bands by using the weights from hidden layers of MPF-NN. MPF derived from MODIS data using MPF-NN model has a RMSE of 3.91\% and a correlation coefficient of 0.73 when compared with validation data, which 
outperforms the previous spectral unmixing algorithm~\cite{rosel2012melt}. It also demonstrates the ability of modeling the melt pond evolution at different sites as well as the Arctic. Finally, though we illustrated the capability of MPF-NN model working with MODIS data, the model can be generalized using various multi-spectral satellite datasets for different parameter retrieval. 

%one produced by the spectral unmixing algorithm in the work of R{\"o}sel et al.~\cite{rosel2012melt}. Furthermore, our MPF-NN model demonstrates a comparable capability in terms of modeling melt pond fraction over the Arctic sea ice, compared with the CICE hindcasts model~\cite{schroder2014september}. As our MPF-NN model has the capability to estimate the MPF trend at an individual site, this provides an opportunity to further study the spatial-temporal variability of melt pond evolution. Though we illustrated the capability of the MPF-NN model with MODIS data, the model can be generalized using various multi-spectral satellite datasets. 

%The model's performance also reflects the effectiveness of using visible imagery as prior knowledge, when considering the spatial-temporal variability. As our future work, we plan to expand our study with more high-quality visible imagery data with larger spatial and temporal coverage. 

\section*{Acknowledgment}
The hign-resolution visible images were provided by NSIDC and MODIS data were obtained from NASA website. This work was supported in part by the National Science Foundation (NSF) under grant No.1251257. 

\ifCLASSOPTIONcaptionsoff
  \newpage
\fi



\bibliographystyle{IEEEtran}
\bibliography{IEEEabrv,refs}
\end{document}